\documentclass[11pt]{article}
\pdfoutput=1
\usepackage[utf8]{inputenc}
\usepackage[margin=1in]{geometry}
\usepackage{amsmath,amssymb,amsthm}
\usepackage{graphicx}
\usepackage{hyperref}
\usepackage{booktabs}
\usepackage{enumitem}
\usepackage{acronym}
\usepackage{orcidlink}    
\usepackage[utf8]{inputenc}
\usepackage[T1]{fontenc}

\newtheorem{definition}{Definition}

\acrodef{MAS}[MAS]{Multi-Agent System}
\acrodef{AI}[AI]{Artificial Intelligence}
\acrodef{BFT}[BFT]{Byzantine Fault Tolerance}
\acrodef{PBFT}[PBFT]{Practical Byzantine Fault Tolerance}
\acrodef{LLM}[LLM]{Language Large Model}
\acrodef{SECP}[SECP]{Self Evolving Coordination Protocol}

\title{Self-Evolving Coordination Protocol in Multi-Agent AI Systems: An Exploratory Systems Feasibility Study}
\date{}
\author{%
  \begin{tabular}{@{}c@{\hspace{1cm}}c@{}} 
    \begin{minipage}[t]{0.42\textwidth}\centering
      Jose Manuel de la Chica Rodriguez \small\orcidlink{0009-0009-9649-5805}\\
      {\small{Head of AI Lab}}\\
      {\small\textit{AI Lab, Grupo Santander}}\\
      {\small\textit{Madrid, Spain}}
    \end{minipage}
    &
    \begin{minipage}[t]{0.42\textwidth}\centering
      Juan Manuel Vera Díaz \small\orcidlink{0000-0002-6152-5789}\thanks{Corresponding author. Email: \texttt{juanma.vera@gruposantander.com}}\\
      {\small{Senior AI Researcher}}\\
      {\small\textit{AI Lab, Grupo Santander}}\\
      {\small\textit{Madrid, Spain}}
    \end{minipage}
  \end{tabular}%
}

\begin{document}

\maketitle

\begin{abstract}
Contemporary multi-agent AI systems increasingly depend on internal coordination mechanisms to combine, arbitrate, or constrain the outputs of heterogeneous components. In safety-critical and regulated domains such as finance. These mechanisms must satisfy strict formal requirements, remain auditable, and operate within clearly bounded limits. Coordination logic therefore functions as a governance layer, not merely an optimization heuristic.

This paper reports an \emph{exploratory systems feasibility study} of \emph{Self Evolving Coordination Protocol} (SECP): decision protocols that permit limited, externally validated self-modification while preserving fixed formal invariants. We study a controlled proof-of-concept setting in which six fixed Byzantine consensus protocol proposals are evaluated by six specialized decision modules (three instantiated with Claude Sonnet~4.5 and three with GPT-4). All coordination regimes operate under identical hard constraints, including Byzantine fault tolerance ($f<n/3$), $O(n^2)$ message complexity, complete non-statistical safety and liveness arguments, and bounded explainability.

Four coordination regimes are compared in a single-shot design: unanimous hard veto, weighted scalar aggregation (control), SECP~v1.0 (an agent-designed non-scalar protocol), and SECP~v2.0 (the result of one governed modification). Outcomes are evaluated using a single metric, \emph{proposal coverage}, defined as the number of proposals accepted by a given protocol. The results reveal a systematic trade-off between coverage and the preservation of non-compensable objection rights. A single recursive modification increased coverage from two to three accepted proposals (a 50\% relative increase) while preserving all declared invariants.

The study makes no claims about statistical significance, optimality, convergence, or general intelligence. Its contribution is architectural rather than predictive: it demonstrates that bounded self-modification of coordination protocols is technically implementable, auditable, and empirically analyzable under explicit formal constraints. This establishes a necessary, though far from sufficient, foundation for evaluating such mechanisms in governed AI systems.
\end{abstract}

\section{Introduction}

\subsection{Motivation: coordination as a governance layer}

Contemporary \ac{AI} deployments in high-stakes domains are rarely monolithic. Instead, they are assemblages of specialized components (formal verifiers, performance optimizers, robustness monitors, cost models), each optimized for different objectives and evaluated by different criteria. System behavior therefore depends not only on the properties of individual components, but critically on the rules and procedures that govern how those components’ judgments are combined. In regulated settings such as finance, healthcare, and safety-critical infrastructure, those governing procedures must satisfy additional constraints: (i) decisions must meet formal safety and liveness requirements, (ii) respect resource and operational limits, and (iii) remain explainable and auditable to human supervisors.

Two simple coordination paradigms frame the problem. Scalar aggregation maps heterogeneous assessments onto a single numeric objective (for example, weighted voting or score averaging). That approach maximizes throughput and makes decision rules mechanically simple, but it collapses structured disagreement. Any concern can be offset by sufficient aggregate support and non-compensable objections disappear. At the opposite extreme, hard-veto schemes preserve full autonomy for each component but routinely produce deadlock when perspectives conflict. Neither extreme is acceptable in many regulated contexts.

This work investigates the middle ground. Coordination mechanisms that retain non-scalar objection rights while enabling rule-based resolution of disagreements. We define and study \acf{SECP}: coordination protocols that permit bounded, auditable modification of their own decision rules according to pre-specified invariants. The intent is not to replace human governance but to provide a formally constrained architectural layer that supports limited adaptation while preserving auditability and safety.

\subsection{Contribution and scope}

The paper offers an empirical, architecture-focused feasibility study rather than a theorem or a statistical evaluation. In a controlled proof-of-concept experiment: six decision modules evaluating six Byzantine consensus proposals. We implemented several coordination regimes (unanimous veto, scalar aggregation, and two \ac{SECP} variants) and executed a single, governed modification of the \ac{SECP}. The study asks a bounded question: can contemporary \ac{AI} systems be arranged to synthesize non-scalar coordination rules and perform a validated one-step protocol revision while maintaining declared invariants?

The experiment yields three concrete observations under those narrow conditions: (i) current models can be used to produce non-scalar coordination logic and to propose a validated parameter modification; (ii) a single validated modification produced a measurable change in the coverage metric while audited invariants held in the observed run; (iii) coordination design exposes a consistent trade-off between coverage (how many proposals are accepted) and evaluator autonomy (the capacity of components to impose non-compensable objections).

For clarity, these are the explicit limits of the paper’s claims:

\subsubsection*{What the paper demonstrates}
\begin{itemize}
    \item Technical feasibility of bounded, governed protocol modification using current \ac{AI} models.
    \item A clear, easily computed coverage metric that captures coordination outcomes.
    \item That protocol modifications can be constrained to preserve declared invariants in a single validated iteration.
    \item The existence of a coverage autonomy trade-off across coordination regimes.
\end{itemize}

\subsubsection*{What the paper does not demonstrate}
\begin{itemize}
    \item Statistical generality or significance.
    \item Convergence, long-run stability, or optimality of the modification process.
    \item Adversarial robustness or resistance to strategic manipulation.
    \item Production readiness or regulatory approval.
    \item That \ac{LLM}-based assessments are a substitute for mechanized proof checking.
\end{itemize}

The experiment is deliberately scoped to be auditable and reproducible. Its purpose is to establish whether a narrowly specified architectural pattern can be implemented and evaluated, not to claim performance, optimality, or deployment suitability.

\subsection{Paper structure}

The remainder of the paper proceeds as follows:

\begin{itemize}
    \item \textbf{Section 2} provides a comprehensive study of the related state-of-the-art
    \item \textbf{Section 3} formalizes the problem setting and defines notation
    \item \textbf{Section 4} describes the experimental methodology and the coordination protocols tested
    \item \textbf{Section 5} reports empirical results.
    \item \textbf{Section 6} interprets those results and discusses theoretical implications.
    \item \textbf{Section 7} examines practical implications for governance in financial systems.
    \item \textbf{Section 8} lists limitations and boundaries of interpretation.
    \item \textbf{Section 9} concludes with recommendations for future work.
\end{itemize}

\section{Related Work}

Distributed systems and \acp{MAS} are the foundational paradigms in modern computing and \ac{AI}. Both areas grapple with the central challenge of coordination among autonomous agents whose local views may be partial, noisy, or adversarial. At its core, consensus ensures a consistent global state between agents, while coordination enables complex tasks through joint behavior. \ac{BFT}, originally formalized through the Byzantine General Problem, provides a model of arbitrary agent failures or malicious behavior. Meanwhile, multi-agent coordination extends consensus to richer agent behaviors such as task allocation, communication dynamics, decentralized learning, and adaptation.

This state-of-the-art survey delineates the foundations, major algorithmic families, scalability and robustness enhancements, and emerging trends at the intersection of Byzantine consensus protocols and multi-agent coordination systems. Crucially, it connects classical results from distributed computing with modern \ac{AI}-driven \ac{MAS} research.

\subsection{Foundations of Byzantine Consensus}


The Byzantine consensus problem was first described in \cite{zhong2023byzantine} as a metaphor to achieve agreement among distributed processes when some nodes may act arbitrarily (“Byzantine faults”). Early solutions, such as the Oral Messages and Signed Messages approaches, established fundamental bounds: consensus in the presence of Byzantine nodes is possible only if $n \geq 3f + 1$, where $n$ is the total number of nodes and $f$ is the number of faulty processes capable of arbitrary behavior including malicious messaging.

\subsubsection*{Classic \ac{BFT} algorithms}
\begin{itemize}
    \item \textbf{\acf{PBFT}}: Introduced in \cite{castro1999practical}, \ac{PBFT} was the first practical protocol for State Machine Replication tolerating Byzantine faults in partially synchronous systems. It uses pre-prepare, prepare, and commit phases to ensure safety and liveness, requiring $3f+1$ replicas to tolerate up to f faulty ones. The algorithm emphasizes cryptographic authentication and view-change mechanisms.
    \item \textbf{HotStuff and Variants}: HotStuff \cite{yin2019hotstuff} simplifies and streamlines \ac{PBFT}’s view-change logic, offering linear-time consensus and pipelining for improved throughput. It has become a basis for several modern consensus frameworks, especially in blockchain systems where performance and modularity matter.
    \item \textbf{Asynchronous \ac{BFT} Protocols}: Protocols like HoneyBadgerBFT \cite{miller2016honey} provide consensus without requiring timing assumptions by using advanced cryptographic primitives and atomic broadcast services to tolerate Byzantine nodes in fully asynchronous environments.
\end{itemize}

\subsection*{Scalability and Multi-leader Approaches}
Scalability has been a long-standing limitation of \ac{PBFT}-like protocols due to their $\mathcal{O}(n^2)$ communication complexity. Recent work has focused on hierarchical and multi-leader designs:

\begin{itemize}
    \item \textbf{Hierarchical and Grouping Protocols}: By creating layers or groups of replicas, systems reduce the number of communication paths needed for consensus. These approaches \cite{thai2019hierarchical, li2020dhbft} elect local leaders and then reconcile group decisions at higher levels, thereby reducing overhead
    \item \textbf{Multi-Leader \ac{BFT}}: Protocols such as BigBFT \cite{salem2021bigbft} and FNFBFT \cite{avarikioti2020fnf} allow multiple leaders to propose blockes or ofering simultaneously, improving throughput and reducing latency compared to single-leader approaches. These designs often achieve communication complexity closer to $\mathcal{O}(n)$ in favorable conditions
    \item \textbf{Threshold Signatures and Dynamic Membership}: Incorporating threshold cryptography and dynamic node join/leave mechanisms (e.g., LTSBFT) improves adaptability and communication efficiency \cite{math12172643}.
\end{itemize}

Overall, the consensus literature has evolved from static, leader-centric algorithms to flexible, layered, and scalable protocols tailored for large, dynamic networks.




\subsection{Multi-Agent Coordination: Consensus Beyond Fault Tolerance}

Consensus in \ac{MAS} abstracts beyond failure tolerance into broader coordination and cooperative behaviors. In contrast to classical distributed consensus, where the goal is consistent ordering or state replication, \ac{MAS} consensus often refers to aligning agent states, opinions, or actions, not just recovering from faults.

\subsubsection*{Consensus Problems in Cooperative Control}

Wei Ren et al. \cite{ren2005survey} provided an influential survey on consensus problems in multi-agent coordination, focusing on cooperative control problems such as formation, synchronization, and state agreement. In these settings, agents exchange local information to converge asymptotically to a shared variable or control law.

The \ac{MAS} consensus literature employs graph theory to model agent communication networks. Agents update their states using weighted averages of neighbors’ states, and protocols are designed to guarantee convergence under both static and time-varying topologies.

\subsubsection*{Coordination Mechanisms in \ac{MAS}}
Multi-agent coordination encompasses various protocol styles:
\begin{itemize}
    \item \textbf{Consensus Protocols for Cooperative Control}: These protocols ensure agents align on common values (e.g., velocity, formation position) through iterative state updates based on neighbors’ states \cite{ren2005survey}.
    \item \textbf{Task Allocation an Negotiation Protocols}: The Contract Net Protocol \cite{smith1988contract} exemplifies how agents negotiate task assignments in decentralized settings, with managers issuing calls for proposals and contractors responding with bids.
    \item \textbf{Decentralized Decision Models}: Frameworks such as Decentralized POMDPs \cite{oliehoek2012decentralized} generalize decision-making under uncertainty and partial observations. They are widely used in coordination problems where communication is limited or noisy.
    \item \textbf{Learning-based Coordination}: Reinforcement learning, graph neural networks, and attention mechanisms are increasingly integrated into \ac{MAS} protocols to improve scalability, adaptability, and robustness to dynamic environments \cite{sun2025multiagentcoordinationdiverseapplications}. These approaches allow agents to learn effective coordination policies rather than relying solely on hand-designed protocols.
\end{itemize}

Emerging research also explores hybrid coordination strategies, unifying hierarchical and decentralized elements or coupling symbolically defined protocols with learned behavior.




\subsection{Integrating Byzantine Consensus into Multi-Agent Coordination}
As \ac{MAS} incorporate autonomous behavior and \ac{AI} components, the role of fault tolerance becomes critical. Byzantine faults in \ac{MAS} can represent adversarial agents, sensor noise, conflicting model outputs (e.g., hallucinations in \ac{LLM} agents), or malicious software infiltrations. Addressing Byzantine challenges within \ac{MAS} calls for integrating consensus resilience into coordination protocols.

\subsubsection*{Byzantine Fault Models for \ac{MAS}}
Traditional \ac{BFT} assumes static nodes with potential misbehavior. However, \ac{MAS} applications, particularly with learning agents, require nuanced fault models that consider emergent behaviors and dynamic roles.

Recent work such as IBGP \cite{mao2024ibgpimperfectbyzantinegenerals} extends the Byzantine Generals framework to practical MAS scenarios, introducing Imperfect Byzantine Generals to account for local coordination requirements without full global consensus, increasing efficiency while maintaining robust local agreement properties.

Systems like the Byzantine fault-tolerant multi-agent healthcare \cite{chadderwala2025byzantinefaulttolerantmultiagenthealthcare} messaging framework demonstrate practical integration of Byzantine consensus into domain-specific \ac{MAS} through gossip protocols and cryptographic validation, illustrating how coordination and fault tolerance can coexist in real-time, distributed \ac{MAS} tasks.

\subsubsection*{Weighted and Confidence-based Consensus}
In cases where agents possess differing levels of reliability (e.g., \ac{LLM}-based agents with confidence metrics), consensus can be weighted to reflect agent trustworthiness. CP-WBFT \cite{zheng2025rethinkingreliabilitymultiagentsystem} introduces confidence probes and weighted Byzantine fault tolerant consensus tailored for high-fault environments, demonstrating applicability in non-traditional \ac{MAS} domains.

This shift reflects a broader trend toward graded trust and reputation mechanisms within consensus algorithms, bridging classical Byzantine fault tolerance with MAS coordination semantics.




\subsection{Open Challenges}
\subsubsection*{Scalability vs Robustness}
A central trade-off in consensus is between scalability (handling large numbers of agents efficiently) and robustness (tolerating adversarial faults). Hierarchical consensus methods \cite{salem2021bigbft} and multi-leader approaches aim to balance these by localizing coordination then reconciling global agreement.

Similarly, in \ac{MAS} consensus, reducing communication overhead (e.g., through partial information exchange or approximation algorithms) must be reconciled with guarantees of global coordination or performance.

\subsubsection*{Asynchrony and Uncertainty}
Asynchronous environments, where message delays, network partitions, and unpredictable timings exist, challenge both Byzantine consensus and \ac{MAS} coordination. Asynchronous \ac{BFT} protocols like HoneyBadgerBFT \cite{ji2024review, miller2016honey} confirm that consensus is achievable without timing assumptions.

For \ac{MAS}, dynamic interaction topologies and decentralized communication intensify these challenges \cite{nguyen2025survey}. Partial observability, network variability, and learning-induced dynamics complicate the design of reliable coordination mechanisms.

\subsubsection*{Emergent and Learned Coordination}
Machine learning, particularly reinforcement learning and neural message passing (e.g., graph neural networks), plays an increasing role in coordination where formal protocols are insufficient \cite{sun2025multiagentcoordinationdiverseapplications}. Such methods allow agents to learn robust policies that adapt to environmental changes and partial observability.

Nonetheless, these approaches introduce new verification challenges: learned behaviors may not adhere to formal safety properties unless explicitly constrained.

\subsubsection*{Unified Frameworks for Robust Coordination}
A crucial research direction is the development of unified frameworks that combine formal Byzantine consensus guarantees with adaptive coordination mechanisms suitable for heterogeneous agent capabilities, dynamic mission objectives, and \ac{AI}-driven decision processes.

\subsubsection*{Trust and Reputation in Consensus}
Incorporating trust, reputation, and agent confidence into consensus protocols reflects real-world scenarios where not all nodes are equal. Weighted consensus, reputational metrics, and confidence probes promise more efficient and fault-aware coordination strategies.

\subsubsection*{Human-\ac{AI} Coordination}
Future \ac{MAS} deployments increasingly involve human agents. Designing consensus and coordination protocols that accommodate human preferences, negotiation, and contextual understanding remains an open area of research, especially where safety and reliability are paramount.

\subsubsection*{Guaranteeing Learned Protocol Safety}
Operationalizing machine learning in consensus protocols challenges existing verification paradigms. Bridging formal methods with empirical learning controllers, perhaps through hybrid symbolic-neural coordination models, is likely to be a major research frontier.

\section{Problem Setting}

\subsection{Byzantine Consensus Algorithm Evaluation}
Byzantine consensus protocols allow distributed systems to agree on a single outcome even when some participants exhibit arbitrary (Byzantine) behavior. The classical feasibility bound states that agreement is possible if and only if the number of faulty nodes $f$ satisfies $f < n/3$, where $n$ is the total number of nodes \cite{lamport1982byzantine,castro1999practical}.

The algorithmic design space for Byzantine consensus is broad and continues to mature. Classical protocols such as \ac{PBFT} \cite{castro1999practical} offer rigorous safety and liveness guarantees under well-defined network assumptions. More recent variants navigate different trade-offs: some prioritize low latency in partially synchronous settings, others maximize throughput under particular failure or workload models, and others emphasize verifiability and implementation simplicity—often at the cost of additional cryptographic machinery or protocol complexity.

In practice, assessing a Byzantine consensus proposal is unavoidably multi-disciplinary. A security specialist focuses on resilience to Byzantine strategies and exploit surfaces; a formal methods specialist scrutinizes safety and liveness arguments; a systems engineer evaluates computational and communication overhead; and an operations specialist weighs explainability, observability, and debuggability. Each perspective is valid, yet they can yield conflicting judgments of the same design: a throughput-optimized protocol may obscure proof structure, while a formally pristine design may impose engineering or operational burdens that undermine deployability.

Reconciling these competing yet legitimate perspectives is precisely the coordination problem addressed in this work.

\subsection{Formal Problem Definition}

\begin{definition}[Proposal Set]
Let
\[
\mathcal{P}=\{C_{EXP},C_{VAL},C_{MIN},G_{ROB},G_{PRF}, G_{ECO}\}.
\]
Each proposal $p\in\mathcal{P}$ is a tuple
\[
p=\big(\Gamma_p,\ \mathrm{Msg}_p,\ \pi^{\mathrm{safe}}_p,\ \pi^{\mathrm{live}}_p,\ x_p\big),
\]
where: (i) $\Gamma_p$ is the fault model specifying tolerated Byzantine failures, complexity bounds on message, time, and space requirements, (ii) $\mathrm{Msg}_p$ message-passing algorithm with defined message types and control flow, (iii) $\pi^{\mathrm{safe}}_p$ and $\pi^{\mathrm{live}}_p$ are proof objects for safety and liveness of $p$ under the proposal's stated fault/network model, and (iv) $x_p$ is a textual operational explanation of $p$.
\end{definition}

The identifiers ($C_{EXP}$, $C_{VAL}$, etc\dots) are generation-time labels and do not constrain evaluation; each $p\in\mathcal{P}$ is assessed solely via its artifacts $(\Gamma_p,\mathrm{Msg}_p,\pi^{\mathrm{safe}}_p,\pi^{\mathrm{live}}_p,x_p)$.

\begin{definition}[Hard Constraints as Predicates]
Fix two semantic predicates:
\[
\mathrm{Safe}(p,n,f)\in\{0,1\},\qquad \mathrm{Live}(p,n,f)\in\{0,1\},
\]
meaning that $\Gamma_p$ with $n$ replicas satisfies safety (resp.\ liveness) against up to $f$ Byzantine replicas under the model stated by $p$.
Fix also:
\[
\mathrm{Check}(p,\pi)\in\{0,1\}\quad\text{(deterministic proof checker)},\qquad \mathrm{wc}(x)\in\mathbb{N}\quad\text{(word count)}.
\]
Define four constraint predicates $H_i:\mathcal{P}\to\{0,1\}$ by:
\begin{itemize}
    \item $H_1(p)$ (Byzantine Tolerance): $\forall n\in\mathbb{N}\ \forall f\in\mathbb{N}\ (3f+1\le n \Rightarrow \mathrm{Safe}(p,n,f)=1 \wedge \mathrm{Live}(p,n,f)=1)$.
    \item $H_2(p)$ (Message Complexity): $\exists c\in\mathbb{R}_{>0}\ \exists n_0\in\mathbb{N}\ \forall n\ge n_0:\ \mathrm{Msg}_p(n)\le c\,n^2$.
    \item $H_3(p)$ (Formal Verification): $\mathrm{Check}(p,\pi^{\mathrm{safe}}_p)=1\ \wedge\ \mathrm{Check}(p,\pi^{\mathrm{live}}_p)=1$.
    \item $H_4(p)$ (Explainability): $\mathrm{wc}(x_p)\le 500$.
\end{itemize}
Let the joint feasibility predicate be $H(p)\;:=\;\bigwedge_{i=1}^4 H_i(p)$ and the feasible proposal set be
\[
\mathcal{P}_H\;:=\;\{p\in\mathcal{P}: H(p)=1\}.
\]
\end{definition}

\textbf{Experimental assumption (pre-screening)}: in the reported experiment, all proposals satisfy the hard constraints, i.e.\ $\mathcal{P}_H=\mathcal{P}$ (equivalently, $\forall p\in\mathcal{P}$, $H(p)=1$).

\begin{definition}[Decision Module]
Let $\mathcal{M}=\{m_1,\dots,m_6\}$ be the fixed set of decision modules. Each module $m\in\mathcal{M}$ is specified by an assessment space $\mathcal{A}_m$ and a fixed evaluation function
\[
\mathrm{eval}_m:\mathcal{P}_H\to\mathcal{A}_m,
\]
which does not adapt during the experiment.
\end{definition}

\textbf{Decision Module Instantiation}: Six modules are instantiated via large language models with system prompts that fix their evaluation rubric:
\begin{itemize}
    \item \textbf{Explorer} (Claude Sonnet 4.5, Anthropic)
    \item \textbf{Validator} (Claude Sonnet 4.5, Anthropic)
    \item \textbf{Minimalist} (Claude Sonnet 4.5, Anthropic)
    \item \textbf{Robustifier} (GPT-4, OpenAI)
    \item \textbf{Proofsmith} (GPT-4, OpenAI)
    \item \textbf{Economizer} (GPT-4, OpenAI)
\end{itemize}

\begin{definition}[Coordination Protocol]
A coordination protocol $\Pi$ is a deterministic aggregation rule
\[
\Pi:\ \prod_{m\in\mathcal{M}} \mathcal{A}_m\ \longrightarrow\ \{\mathrm{Accept},\mathrm{Reject}\}.
\]
For each proposal $p\in\mathcal{P}_H$, define the induced decision
\[
\mathrm{dec}_\Pi(p)\;:=\;\Pi\big((\mathrm{eval}_m(p))_{m\in\mathcal{M}}\big).
\]
\end{definition}

Coordination protocols vary only in $\Pi$; the sets $\mathcal{P}$ and $\mathcal{M}$, the predicates $H_1$--$H_4$, and all $\mathrm{eval}_m$ are fixed across conditions.

\begin{definition}[Protocol Coverage]
For a coordination protocol $\Pi$, define coverage over feasible proposals by
\[
\Delta S(\Pi)\;:=\;\big|\{p\in\mathcal{P}_H:\ \mathrm{dec}_\Pi(p)=\mathrm{Accept}\}\big|.
\]
\end{definition}

Since $\mathcal{P}_H=\mathcal{P}$ in our experiment and $|\mathcal{P}|=6$, we have $\Delta S(\Pi)\in\{0,1,2,3,4,5,6\}$. Coverage is deliberately a raw count (no utility weights, ranking, or statistical aggregation), isolating the effect of the aggregation rule $\Pi$.

\subsection{Recursive Protocol Improvement: Architectural Capacity vs. Convergence}

\begin{definition}[Recursive Modification Capability]
A coordination system has \emph{recursive modification capability} if there exist:
(i) per-module suggestion functions $\mathrm{mod}_m$ into a space of edits $\mathcal{U}$,
(ii) a constructor $\mathrm{Revise}$ producing a candidate protocol from the current one and the edit profile, and
(iii) an invariant predicate $\mathrm{Inv}$ over coordination protocols,
such that for every version $v\in\mathbb{N}$ the next version is defined by
\[
\Pi_{v+1}\;:=\;
\begin{cases}
\mathrm{Revise}\big(\Pi_v,\ (\mathrm{mod}_m(\Pi_v))_{m\in\mathcal{M}}\big), & \text{if }\mathrm{Inv}\!\left(\mathrm{Revise}(\Pi_v,(\mathrm{mod}_m(\Pi_v))_{m\in\mathcal{M}})\right)=1,\\
\Pi_v, & \text{otherwise (rollback)}.
\end{cases}
\]
\end{definition}

\begin{definition}[Protocol Improvement Iteration]
Given a versioned coordination protocol sequence $(\Pi_v)_{v\in\mathbb{N}}$ as above, an \emph{improvement iteration} occurs at step $v\to v+1$ when:
\begin{enumerate}
    \item $\Pi_{v+1}\neq \Pi_v$ (a non-trivial revision is adopted, hence $\mathrm{Inv}(\Pi_{v+1})=1$ by construction), and
    \item coverage is measured for both versions and compared, i.e.\ $\Delta S(\Pi_{v+1})$ versus $\Delta S(\Pi_v)$.
\end{enumerate}
\end{definition}

\textbf{Critical terminological clarification}: we use ``recursive'' to denote the \emph{capacity} encoded by $(\mathrm{mod}_m,\mathrm{Revise},\mathrm{Inv})$ to repeatedly produce validated protocol revisions. Empirically, the paper executes exactly one improvement step (\acs{SECP} v1.0 $\to$ \acs{SECP} v2.0); it does not claim multi-step convergence, optimality, or long-run stability, which would require many iterations beyond this proof-of-concept.

\begin{figure}[t]
    \centering
    \includegraphics[width=\textwidth]{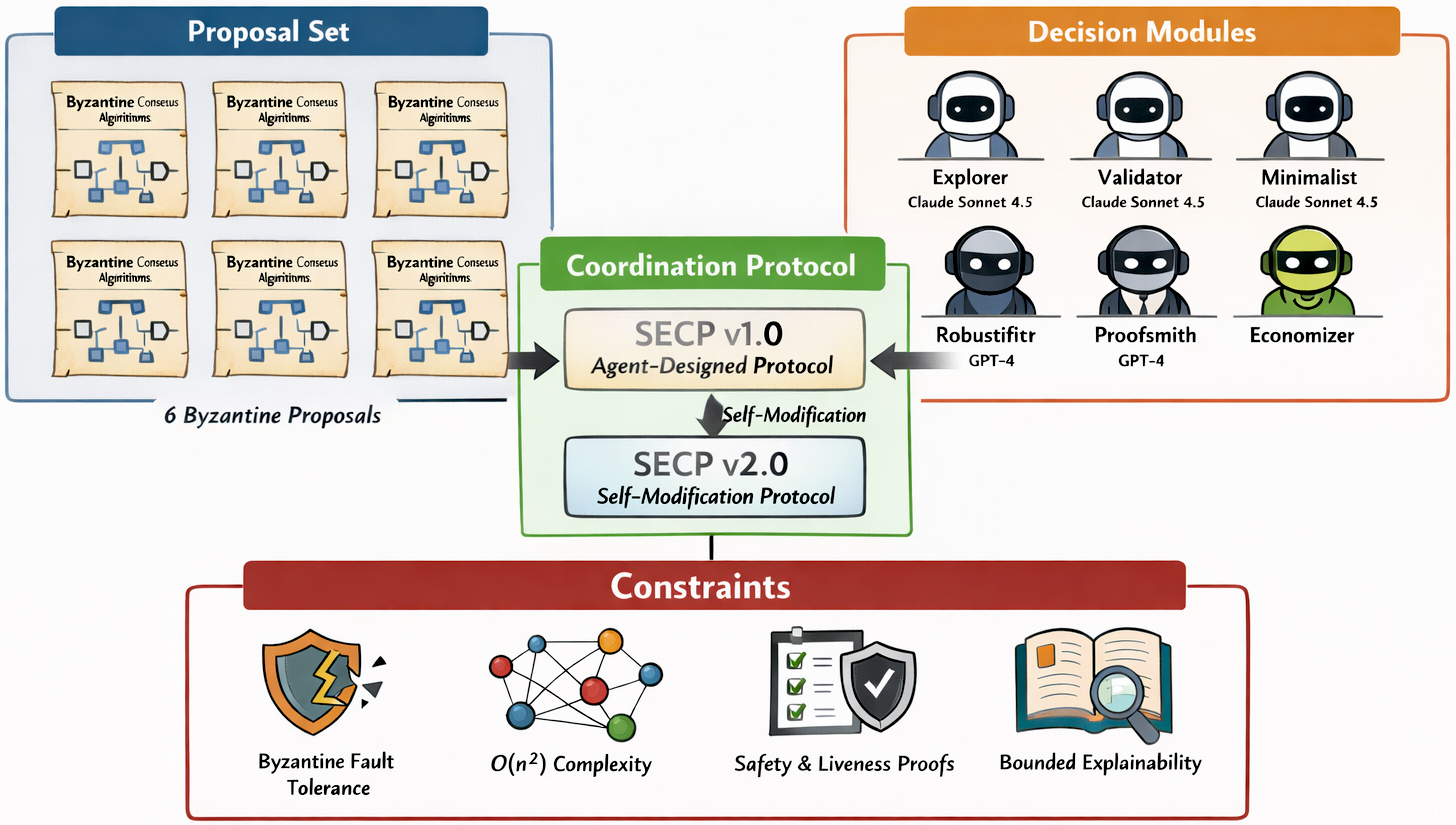}
    \caption{Schematic overview of the problem setting. A fixed set of Byzantine consensus proposals is evaluated by a heterogeneous set of decision modules. Their assessments are aggregated by a coordination protocol operating under fixed hard constraints (Byzantine fault tolerance, complexity bounds, formal correctness, and bounded explainability). The coordination protocol itself may undergo bounded, validated modification while preserving these invariants.}
    \label{fig:problem-setting}
\end{figure}

To support the formal definitions introduced in this section, Figure \ref{fig:problem-setting} provides a schematic overview of the problem setting. The diagram summarizes the fixed proposal set, the heterogeneous decision modules, the coordination protocol, and the hard constraints under which all evaluations are performed. Its purpose is not to introduce new concepts, but to offer a structural view of how the formally defined components interact within the experimental system.

\section{Methodology}

\subsection{Experimental Design: Single-Shot Feasibility Study}

\textbf{Design rationale.} We use a \emph{single-shot} design: each coordination protocol is executed exactly once on the same fixed proposal set $\mathcal{P}$ and the same fixed module set $\mathcal{M}$. The objective is architectural \emph{feasibility} and interpretability, not statistical estimation.

\subsubsection*{Why single-shot is appropriate}
\begin{enumerate}
    \item \textbf{Proof-of-concept objective.} The primary question is \emph{existence/implementability}: can a governed, bounded self-modification loop for coordination be implemented with current AI systems?
    \item \textbf{Protocol determinism.} For a fixed set of module outputs, each coordination protocol $\Pi$ is deterministic. Repeating trials would primarily probe \ac{LLM} output stochasticity rather than the coordination logic itself.
    \item \textbf{Oversight cost.} Each end-to-end run requires non-trivial external oversight (invariant checks, log review, and safety validation). Repeated trials would scale linearly in effort.
    \item \textbf{Architectural vs.\ statistical claims.} The contribution is an auditable coordination \emph{architecture} (including a modification and validation workflow), not a performance claim requiring confidence intervals.
\end{enumerate}

\subsubsection*{Acknowledged limitation}
Single-shot results do not support significance testing, confidence intervals, or claims about expected performance under repeated sampling. They should be interpreted as \emph{existence proofs} (an outcome \emph{can} occur) rather than \emph{distributional claims} (an outcome \emph{typically} occurs). We discuss this limitation explicitly in Section \ref{sec:limitations}.

\subsection{Evaluation Interface and Procedure}

\subsubsection*{Module output interface}
For each feasible proposal $p\in\mathcal{P}_H$ (i.e., $H(p)=1$), each module $m\in\mathcal{M}$ produces an assessment record
\[
a_m(p) \;=\; \big(s_m(p),\ r_m(p),\ O_m(p)\big),
\]
where:
\begin{itemize}
    \item $s_m(p)\in[0,1]$ is a deterministic score computed from a structured questionnaire (Appendix~B),
    \item $r_m(p)\in\{\mathrm{Accept},\mathrm{Veto}\}$ is a binary recommendation derived from the same assessment rubric,
    \item $O_m(p)$ is a (possibly empty) set of \emph{constructive objections}, each of which must state a concrete technical deficiency with a verifiable reference to the proposal (e.g., a missing case in a proof, an unhandled network behavior, or an incomplete complexity argument).
\end{itemize}

\subsubsection*{Evaluation pipeline}
Each protocol $\Pi$ is evaluated independently under identical conditions:
\begin{enumerate}
    \item \textbf{Hard-constraint gate.} Compute $\mathcal{P}_H=\{p\in\mathcal{P}:H(p)=1\}$ by verifying $H_1$--$H_4$ externally. Any $p\notin\mathcal{P}_H$ is rejected without invoking $\Pi$.
    \item \textbf{Module assessments.} For every $p\in\mathcal{P}_H$ and $m\in\mathcal{M}$, produce $a_m(p)$.
    \item \textbf{Protocol execution.} For each $p\in\mathcal{P}_H$, the protocol computes a decision:
\[
\mathrm{dec}_\Pi(p)\in\{\mathrm{Accept},\mathrm{Reject}\}
\quad\text{as a deterministic function of }(a_m(p))_{m\in\mathcal{M}}.
\]
    \item \textbf{Coverage.} Compute coverage $\Delta S(\Pi)=|\{p\in\mathcal{P}_H:\mathrm{dec}_\Pi(p)=\mathrm{Accept}\}|$.
    \item \textbf{Audit log.} Persist complete transcripts: $(a_m(p))_{m,p}$, all intermediate protocol states (if any), the final decisions, and a decision rationale trace sufficient for post-hoc review.
\end{enumerate}

No training, re-sampling, repeated trials, or statistical inference is performed. The design isolates the effect of the coordination logic under fixed inputs and fixed feasibility constraints.

\subsection{Phase 1 Baseline: Hard-Veto Unanimity}

Phase~1 implements the minimal coordination rule: unanimous acceptance with mutual veto.

\subsubsection*{Decision rule}
For each $p\in\mathcal{P}_H$, accept if no module vetoes:
\[
\mathrm{dec}_{\Pi}^{\mathrm{P1}}(p)=
\begin{cases}
\mathrm{Accept}, & \text{if } r_m(p)=\mathrm{Accept}\ \ \forall m\in\mathcal{M},\\
\mathrm{Reject}, & \text{otherwise.}
\end{cases}
\]

\subsubsection*{Characteristics}
There is no deliberation or disagreement resolution beyond unanimity. Any single veto implies rejection. This provides a lower-bound baseline on coverage under maximal module autonomy.

\subsection{Control: Weighted Scalar Aggregation}

The control condition implements coordination via compensatory scalar optimization.

\subsubsection*{Scoring}
Each module produces $s_m(p)\in[0,1]$ from a deterministic questionnaire mapping item responses $\{\mathrm{No},\mathrm{Partial},\mathrm{Yes}\}$ to $\{0,0.5,1\}$ and aggregating via a fixed per-module weighting scheme (Appendix~\ref{sec:appendixb}).

\subsubsection*{Aggregation and decision rule}
Define the global score as an equal-weight average:
\[
\mathrm{S}_{\mathrm{Global}}(p) \;=\; \frac{1}{|\mathcal{M}|}\sum_{m\in\mathcal{M}} s_m(p).
\]
Then
\[
\mathrm{dec}_{\Pi}^{\mathrm{ctrl}}(p)=
\begin{cases}
\mathrm{Accept}, & \text{if }\mathrm{S}_{\mathrm{Global}}(p) \ge \tau,\\
\mathrm{Reject}, & \text{otherwise,}
\end{cases}
\]

where $\tau$ is a threshold value empirically fixed by default to $0.6$.

\subsubsection*{Characteristics}
Objections are fully compensable: low scores from one module can be offset by high scores from others. This provides an upper-bound reference for coverage under predetermined, fully compensatory coordination rules.

\subsection{\acs{SECP} v1.0: Agent-Designed Non-Scalar Coordination}

After observing Phase~1 deadlock (universal rejection despite $H(p)=1$ for all $p\in\mathcal{P}$), modules were tasked with jointly designing a coordination protocol that (i) increases coverage relative to $\Pi^{\mathrm{P1}}$, (ii) preserves the hard-constraint gate, (iii) does not reduce to a single scalar threshold, (iv) provides principled disagreement handling, and (v) terminates in bounded time.

\acs{SECP} v1.0 consists of five components applied in sequence.

\subsubsection*{Component 1: Pareto filtering}
Let the score vector be $s(p)=\{s_m(p)\}_{m\in\mathcal{M}}\in[0,1]$.
Define Pareto dominance: $p\prec q$ if
\[
\forall m\in\mathcal{M},\ s_m(q)\ge s_m(p)\quad\text{and}\quad \exists m\in\mathcal{M},\ s_m(q)>s_m(p).
\]
Any Pareto-dominated $p$ (i.e., $\exists q\in\mathcal{P}_H$ with $p\prec q$) is rejected.

\subsubsection*{Component 2: Minimax-regret screening}
Define per-module regret relative to the best observed proposal under that module:
\[
\mathrm{R}_m(p)\;=\;\max_{q\in\mathcal{P}_H}s_m(q)\;-\;s_m(p),
\qquad
R_{\max}(p)\;=\;\max_{m\in\mathcal{M}}\mathrm{R}_m(p).
\]
Reject if $R_{\max}(p)>\rho$, for a fixed regret bound $\rho\in[0,1]$ chosen as a protocol parameter. This enforces a fairness constraint: proposals that impose large regret on \emph{any} module are filtered even if their average score is high.

\subsubsection*{Component 3: Multi-round deliberation with adaptive majority.}
Deliberation proceeds for rounds $t=1,\dots,T_{\max}$ with a pre-specified schedule of support thresholds $(\theta_t,\kappa_t)$ where $\theta_t\in[0,1]$ and $\kappa_t\in\{1,\dots,|\mathcal{M}|\}$.
Define the supporter set in round $t$ as:
\[
\mathrm{S}_t(p)=\{m\in\mathcal{M}: s_m(p)\ge \theta_t\}.
\]
A proposal qualifies on support in round $t$ if $|\mathrm{S}_t(p)|\ge \kappa_t$. The schedule is chosen so that the support requirement relaxes across rounds (i.e., $\kappa_t$ is non-increasing and/or $\theta_t$ is non-increasing with $t$).

\subsubsection*{Component 4: Constructive objections requirement}
Only \emph{constructive} objections are decision-relevant. Formally, an objection in $O_m(p)$ must specify a concrete, checkable technical issue. Non-constructive disagreement (e.g., ungrounded preference) is ignored for voting purposes. A proposal can be accepted only if there are no \emph{unresolved} constructive objections at the acceptance round (i.e., every objection has been explicitly withdrawn or addressed by a documented technical rebuttal acknowledged by the objecting module).

\subsubsection*{Component 5: Bounded termination}
A finite round limit $T_{\max}$ is enforced. If no round yields acceptance, the proposal is rejected.

\subsubsection*{Acceptance criterion (\acs{SECP} v1.0).}
A proposal $p\in\mathcal{P}_H$ is accepted by \acs{SECP} v1.0 if:
\begin{enumerate}
    \item $p$ is not Pareto-dominated;
    \item $R_{\max}(p)\le \rho$;
    \item there exists $t\le T_{\max}$ such that $|\mathrm{S}_t(p)|\ge \kappa_t$; and
    \item no unresolved constructive objections remain at that round.
\end{enumerate}

\subsection{\acs{SECP} v2.0: One Iteration of Governed Protocol Modification}

After evaluating \acs{SECP} v1.0, modules proposed parameter-level modifications (e.g., regret bound $\rho$, the schedule $(\theta_t,\kappa_t)$, and tie-breaking details). The five-component structure was preserved.

\subsubsection*{Modification validation workflow}
A candidate revision from version $v$ to $v+1$ is adopted only if:
\begin{enumerate}
    \item \textbf{Invariant checks pass:} the revision preserves the required invariants stated below (including bounded termination and non-scalar structure).
    \item \textbf{Supermajority approval:} at least $4$ of $6$ modules approve the change.
    \item \textbf{Documentation:} the change rationale, parameter deltas, and expected effect are recorded.
    \item \textbf{Versioning/rollback:} the prior version remains available for rollback if validation fails or regressions are observed.
\end{enumerate}

\subsubsection*{Acceptance criterion (\acs{SECP} v2.0)}
\acs{SECP} v2.0 applies the same decision structure as \acs{SECP} v1.0, but with the modified (validated) parameter values.

\subsection{Safety and Governance Invariants}

Across all conditions, the experiment enforces the following invariants:

\begin{itemize}
    \item \textbf{Feasibility gating:} no protocol may accept $p\notin\mathcal{P}_H$ (i.e., acceptance is only defined over proposals that pass $H_1$--$H_4$).
    \item \textbf{Auditability:} every decision records the protocol identifier/version, the participating module assessments $(a_m(p))_{m\in\mathcal{M}}$, and a rationale trace sufficient for external review.
    \item \textbf{Termination:} every protocol returns $\mathrm{dec}_\Pi(p)\in\{\mathrm{Accept},\mathrm{Reject}\}$ in finite time; in deliberative protocols this is enforced by a finite round bound $T_{\max}$.
    \item \textbf{Version control:} any protocol change is versioned, documented, and reversible (rollback).
\end{itemize}

Invariant preservation is checked empirically through enforced logging, external verification steps, and audit review. This feasibility study does not provide machine-checked formal proofs that the invariants hold for all executions; providing such proofs is deferred to future work requiring additional formalization of the coordination and modification machinery.

\section{Results}

Table~\ref{tab:results} summarizes the coverage achieved by each coordination protocol on the fixed proposal set $\mathcal{P}$. Recall that coverage is
\[
\Delta S(\Pi)=\big|\{p\in\mathcal{P}_H:\mathrm{dec}_\Pi(p)=\mathrm{Accept}\}\big|,
\]
and that all proposals listed as accepted were independently verified to satisfy the hard constraints ($\mathcal{P}_H=\mathcal{P}$ in this experiment).

\begin{table}[h]
\centering
\caption{Protocol coverage on a fixed Byzantine consensus proposal set. $\Delta S(\Pi)$ is the number of accepted proposals from $\mathcal{P}=\{\text{C}_\text{EXP},\text{C}_\text{VAL},\text{C}_\text{MIN},\text{G}_\text{ROB},\text{G}_\text{PRF},\text{G}_\text{ECO}\}$. All accepted proposals satisfy the hard constraints.}
\label{tab:results}
\begin{tabular}{@{}lcl@{}}
\toprule
\textbf{Protocol} & \textbf{$\Delta S(\Pi)$} & \textbf{Accepted Proposals} \\
\midrule
Phase 1: Hard Veto Unanimity & 0 & (none) \\
Control: Weighted Scalar Aggregation & 6 & all six proposals \\
\acs{SECP} v1.0 (non-scalar) & 2 & $\text{C}_\text{VAL}$, $\text{G}_\text{PRF}$ \\
\acs{SECP} v2.0 (one modification) & 3 & $\text{C}_\text{VAL}$, $\text{G}_\text{PRF}$, $\text{G}_\text{ROB}$ \\
\bottomrule
\end{tabular}
\end{table}

\subsection{Phase 1: Universal Rejection Under Unanimity}

Phase~1 produced $\Delta S(\Pi_{\mathrm{P1}})=0$. This outcome occurred despite the fact that every proposal in $\mathcal{P}$ satisfied the hard constraints. In each case, at least one module issued a veto grounded in its specialized evaluation perspective.

Qualitative inspection of the logged objections indicates a consistent pattern of \emph{cross-criterion conflict}:
\begin{itemize}
    \item Proposals emphasizing efficiency attracted objections from proof-focused modules regarding the completeness or structure of the formal arguments.
    \item Proposals emphasizing proof clarity attracted objections from efficiency-focused modules regarding computational or communication overhead.
    \item Proposals introducing novel mechanisms attracted objections from simplicity-focused modules regarding implementability and operational complexity.
\end{itemize}

Because Phase~1 provides no mechanism to reconcile trade-offs, unanimity requires simultaneous satisfaction of all perspectives. The observed deadlock therefore illustrates a basic coordination failure mode: even when all candidates are formally feasible, a multi-perspective evaluation can yield systematic rejection in the absence of disagreement-resolution structure.

\subsection{Control: Universal Acceptance via Scalar Compensation}

The control protocol achieved $\Delta S(\Pi_{\mathrm{ctrl}})=6$: all proposals were accepted under equal-weight scalar aggregation with threshold $\tau=0.6$. In this condition, each proposal's average score exceeded the acceptance threshold, so no single module's low score was decisive.

This result highlights the defining property of compensatory aggregation: objections are \emph{fully offsettable}. Converting a multi-dimensional assessment profile into a single scalar enables acceptance decisions based on aggregate support rather than perspective-wise agreement, thereby maximizing coverage under fixed, predetermined rules.

\subsection{\acs{SECP} v1.0: Selective Acceptance with Non-Scalar Reasoning}

\acs{SECP} v1.0 achieved $\Delta S(\Pi_{\mathrm{SECP1}})=2$, accepting two proposals: $\text{C}_\text{VAL}$ and $\text{G}_\text{PRF}$. Both accepted proposals satisfied the full \acs{SECP} v1.0 decision pipeline: they were not Pareto-dominated, their minimax regret fell within the configured bound, they obtained sufficient multi-module support within the bounded deliberation schedule, and all recorded constructive objections were resolved at the acceptance stage.

The remaining four proposals were rejected for one or more structurally explicit reasons encoded by \acs{SECP} v1.0 (e.g., Pareto dominance by an alternative, excessive minimax regret for at least one module, or insufficient alignment within the finite deliberation budget). Importantly, these rejections are not attributable to hard-constraint failure; rather, they reflect the protocol's non-scalar selectivity under multi-perspective disagreement.

Overall, \acs{SECP} v1.0 yields intermediate coverage between Phase~1 (deadlock) and the control (universal acceptance), demonstrating that non-scalar coordination can escape unanimity failure while still enforcing non-compensatory safeguards.

\subsection{\acs{SECP} v2.0: Coverage Increase After One Governed Modification}

\acs{SECP} v2.0 achieved $\Delta S(\Pi_{\mathrm{SECP2}})=3$, accepting $\text{C}_\text{VAL}$, $\text{G}_\text{PRF}$, and additionally $\text{G}_\text{ROB}$. The transition from v1.0 to v2.0 consisted of parameter-level adjustments (e.g., regret bound and deliberation schedule) proposed by modules and adopted via the specified validation workflow, including super-majority approval (5 of 6 modules) and external invariant checks.

Under the refined parameters, the two proposals accepted in v1.0 remained accepted, and $\text{G}_text{ROB}$ crossed the acceptance boundary while still satisfying the hard constraints. 

\[
\Delta S(\Pi)[\%] = \frac{\Delta S(\Pi_{\mathrm{SECP2}})-\Delta S(\Pi_{\mathrm{SECP1}})}{\Delta S(\Pi_{\mathrm{SECP1}})} = 50\%
\]

For clarity, the observed coverage progression is:
\begin{itemize}
    \item Phase~1 $\rightarrow$ \acs{SECP} v1.0: $0 \rightarrow 2$ (escape from unanimity deadlock),
    \item \acs{SECP} v1.0 $\rightarrow$ \acs{SECP} v2.0: $2 \rightarrow 3$ (gain from one governed modification),
    \item \acs{SECP} v2.0 versus control: $3$ versus $6$ (non-scalar coordination remains intentionally selective).
\end{itemize}

\subsection{Interpretation: Feasibility of Bounded Self-Modification}

Taken together, the results support the paper's feasibility claim: \textit{\textbf{bounded, governed protocol self-modification can be implemented and can yield measurable changes in behavior under fixed experimental inputs}}. Concretely, the run demonstrates:
\begin{enumerate}
    \item \textbf{Implementability:} modules designed a non-scalar coordination protocol (\acs{SECP} v1.0) after observing a baseline deadlock.
    \item \textbf{Modification capability:} modules proposed and adopted a validated revision (v1.0 $\rightarrow$ v2.0).
    \item \textbf{Measurable impact:} the revision increased coverage ($50\%$ relative coverage improvement);
    \item \textbf{Invariant preservation:} accepted proposals remained within the hard-constraint feasible set and decisions remained auditable;
    \item \textbf{Bounded behavior:} all protocols terminated with a definite decision for every $p\in\mathcal{P}$ within the specified bounds.
\end{enumerate}

As in the methodology, these findings should be interpreted as \emph{existence results from a single execution}. They do not constitute statistical claims about expected coverage under repeated trials, nor do they establish convergence or optimality of the modification process.

\section{Discussion}

\subsection{The Coverage: Autonomy Trade-off}

The results expose a clear and systematic trade-off between \emph{coverage} (the ability of a coordination protocol to accept proposals) and \emph{decision autonomy} (the extent to which individual modules can enforce non-compensable objections).

\subsubsection*{Extreme I: Full Autonomy, Zero Coverage (Phase~1)}  
Under unanimous veto, each module applies its criteria independently and can block any proposal. This preserves maximal autonomy but produces complete deadlock: no proposal simultaneously satisfies all specialized perspectives. The result is $\Delta S=0$, despite all proposals meeting the hard constraints. This outcome is not anomalous; it is the expected consequence of uncoordinated multi-perspective evaluation.

\subsubsection*{Extreme II: Full Coverage, Minimal Autonomy (Control)}  
Scalar aggregation eliminates deadlock by making all objections compensable. Individual assessments influence outcomes only through fixed weights and thresholds; no module can enforce a hard objection. This guarantees maximal coverage ($\Delta S=6$) but at the cost of collapsing structured disagreement into a single scalar. The protocol decides efficiently, but module autonomy is substantially reduced.

\subsubsection*{Intermediate Regimes: Structured Coordination (\acs{SECP} v1.0, v2.0)}  
The \ac{SECP} protocols occupy intermediate positions. Modules retain meaningful blocking power through non-compensable criteria (Pareto dominance, regret bounds, unresolved constructive objections), yet coordination rules allow acceptance when sufficient structured support exists. Coverage is neither minimal nor maximal ($\Delta S=2$ and $3$ respectively), reflecting deliberate selectivity rather than failure. These regimes demonstrate that coordination can relax unanimity without fully suppressing dissent.

\subsection{Recursive Modification: Architectural Capacity Without Convergence Claims}

The transition from \ac{SECP} v1.0 to \ac{SECP} v2.0 shows that a system endowed with recursive modification capability can execute a governed protocol revision with observable effects.

\subsubsection*{What was demonstrated}  
A single modification iteration that:
\begin{enumerate}
    \item was initiated by modules based on analysis of prior outcomes,
    \item preserved all declared invariants via an explicit validation procedure,
    \item increased coverage from $2$ to $3$ accepted proposals, and
    \item retained non-scalar decision logic and bounded termination.
\end{enumerate}

\subsubsection*{What was not demonstrated}  
No evidence is provided for multi-step convergence, optimality, long-term stability, robustness across proposal distributions, or monotonic improvement. The experiment executes exactly one recursive step. Accordingly, the term \emph{recursive} refers to architectural \emph{capacity} for repetition, not to empirically observed recursive dynamics.

This distinction is substantive: \emph{\textbf{the work establishes that bounded self-modification is feasible to implement and govern, not that it reliably improves outcomes or converges under iteration}}.

\subsection{Validity Considerations: \ac{LLM} as Evaluators}

\subsubsection*{Limits of \ac{LLM}-Based Judgment}

Decision modules are instantiated using \acp{LLM} that assess technical artifacts, including formal proofs. This introduces non-trivial validity risks.

\begin{itemize}[leftmargin=*]
    \item[] \textbf{Known limitations}
    \begin{itemize}[label={\labelitemi}]
        \item \textbf{Hallucination}: plausible but incorrect claims about correctness or completeness,
        \item \textbf{Lack of formal verification}: inability to mechanically check proofs,
        \item \textbf{Prompt sensitivity}: assessments may vary with formulation,
        \item \textbf{Reasoning inconsistency}: no guarantee of global coherence across evaluations.
    \end{itemize}
    \item[] \textbf{Mitigation measures applied}
    \begin{enumerate}[label={\labelitemi}]
        \item Hard constraints ($H_1$--$H_4$) are verified independently of module judgments.
        \item Module scoring is constrained by structured, deterministic questionnaires.
        \item Multiple specialized modules provide partial redundancy.
        \item \ac{SECP} protocols preserve non-compensable objection mechanisms, limiting the impact of isolated acceptance errors.
    \end{enumerate}
    \item[] \textbf{Residual risk}
    \begin{itemize}[leftmargin=*]
        \item[] These mitigation measures reduce but do not eliminate error. The experiment evaluates coordination behavior \emph{conditional on module outputs}. It does not validate the correctness of those outputs. Replacing or augmenting \ac{LLM} judgment with formal proof assistants remains necessary for high-assurance deployment.
    \end{itemize}
\end{itemize}

\subsubsection*{Coordination Study. Not an Evaluation Benchmark}

An alternative and arguably more accurate interpretation is that this work studies \emph{coordination among evaluators}, not the objective quality of consensus protocols:
\begin{itemize}
    \item Proposals serve as controlled stimuli to induce disagreement,
    \item Module outputs represent perspective-dependent judgments,
    \item Coordination protocols are the object of analysis,
    \item Coverage measures disagreement resolution capacity, not proposal merit.
\end{itemize}

Under this framing, the technical correctness of any specific acceptance is secondary to the behavior of the coordination mechanism itself.

\subsection{Implications for Coordination Protocol Design}

The observed trade-off implies that no single coordination regime is universally optimal. Design choices depend on institutional priorities.

\begin{itemize}
    \item \textbf{Safety-critical contexts favor low coverage and high autonomy}: false positives dominate false negatives. Unanimity or near-unanimity regimes are defensible despite deadlock risk.
    \item \textbf{Exploratory or high-throughput contexts favor high coverage}: rejecting viable options is costly. Scalar or weakly non-scalar aggregation is acceptable despite reduced autonomy.
    \item \textbf{Regulated financial systems sit between these extremes}. Both false positives (deploying flawed systems) and false negatives (blocking innovation) carry material cost. Intermediate, auditable coordination regimes such as \ac{SECP}-style protocols are therefore more appropriate.
\end{itemize}

The recursive modification capability demonstrated by \ac{SECP} v2.0 suggests that systems need not fix their position on this spectrum ex-ante. However, any adaptation must remain tightly bounded by invariants defining unacceptable behavior.

\subsection{Limits of a Single-Iteration Demonstration}

A single modification step provides only minimal insight into system dynamics. The following remain unresolved:
\begin{itemize}[]
    \item whether further iterations would improve, plateau, oscillate, or degrade performance;
    \item whether convergence to a stable protocol is possible or desirable;
    \item whether non-scalar coordination can approach scalar coverage levels under iteration;
    \item whether the observed $2\rightarrow3$ improvement is typical or incidental.
\end{itemize}

Answering these questions requires longitudinal experimentation with multiple modification cycles. This study does not attempt such analysis.

\section{Practical Implications for Financial AI Governance}

\subsection{Relevance to Financial Systems}

Although the experimental setting is Byzantine consensus, the coordination problem studied here closely mirrors that of financial \ac{AI} systems, where multiple specialized models must jointly determine actions under hard regulatory and risk constraints. In such systems, decisions are rarely produced by a single model; instead, they emerge from the interaction of components with distinct objectives, assumptions, and failure modes. The coordination mechanisms, rather than the domain-specific logic of any single component, often determine whether the system behaves safely, efficiently, and audibly.

\subsubsection*{Algorithmic Trading}

Modern algorithmic trading platforms integrate execution optimizers, market impact estimators, real-time risk controls, and compliance filters. These components frequently disagree. Strategies that minimize execution cost may increase risk exposure or trigger compliance concerns.

Current coordination approaches exhibit clear weaknesses. Fixed veto hierarchies are rigid and prone to deadlock, while scalar optimization frameworks obscure the origin of decisions and are difficult to justify to regulators. A \ac{SECP}-style coordination regime offers a middle ground. Regulatory and risk limits can be enforced as immutable invariants, ensuring non-negotiable compliance. Within those bounds, structured coordination allows explicit trade-offs among performance objectives, with parameter adjustments reflecting changing market regimes. Crucially, every decision and protocol change is logged, supporting post-trade analysis and regulatory review.

\subsubsection*{Credit Decisioning}

Credit decisioning systems combine default risk estimation, profitability analysis, fairness constraints, and portfolio-level considerations. These perspectives are often in tension, and fixed scorecards or monolithic models struggle to adapt to evolving economic conditions without violating regulatory expectations.

In this context, \ac{SECP}-style coordination supports a clear separation between what is fixed and what is adaptable. Fair-lending rules and regulatory requirements function as invariants that cannot be overridden. Within this constrained space, coordination protocols manage risk/return trade-offs and can adapt parameters as macroeconomic conditions or portfolio composition change. Because decisions remain decomposable into module-level assessments and coordination logic, the system can generate explicit rationales suitable for adverse action notices and supervisory scrutiny.

\subsubsection*{Fraud Detection}

Fraud detection systems must balance false positives, which impose customer friction, against false negatives, which generate direct financial loss. Multiple detection models frequently produce conflicting signals, and naive aggregation can either over-block legitimate customers or under-detect sophisticated fraud.

A \ac{SECP}-style approach enforces customer protection and regulatory compliance as non-compensable constraints while allowing graded responses based on coordinated confidence levels. As fraud patterns evolve, coordination parameters (such as response thresholds) can be adjusted in a controlled manner, without weakening core protections. The resulting decisions remain auditable and explainable, which is essential for customer appeals and regulatory review.

\subsection{Requirements for Production Deployment}

Deploying recursively modifiable coordination protocols in production financial systems requires governance structures that go beyond this feasibility study. In practice, a layered architecture is necessary. At the base sits an invariant layer containing immutable regulatory, legal, and risk constraints. Above it, a protocol layer defines coordination logic that may be modified, but only within clearly specified bounds. Operational components execute decisions within these constraints, while an audit layer records all assessments, decisions, protocol versions, and modifications.

Human oversight remains essential. Significant protocol changes must require explicit approval by designated authorities, continuous monitoring must track coverage and error signals, and rollback mechanisms must allow rapid reversion if degradation is detected. These controls are not optional; without them, recursive modification introduces unacceptable operational and regulatory risk.

Risk management must also address longer-term dynamics. Drift detection is required to identify gradual deviations from intended behavior, adversarial testing is necessary to prevent strategic manipulation, and explicit limits must constrain the growth of protocol complexity over successive modifications. Continuous verification against regulatory requirements is mandatory, particularly as regulations evolve.

\begin{figure}[t]
    \centering
    \includegraphics[width=0.6\textwidth]{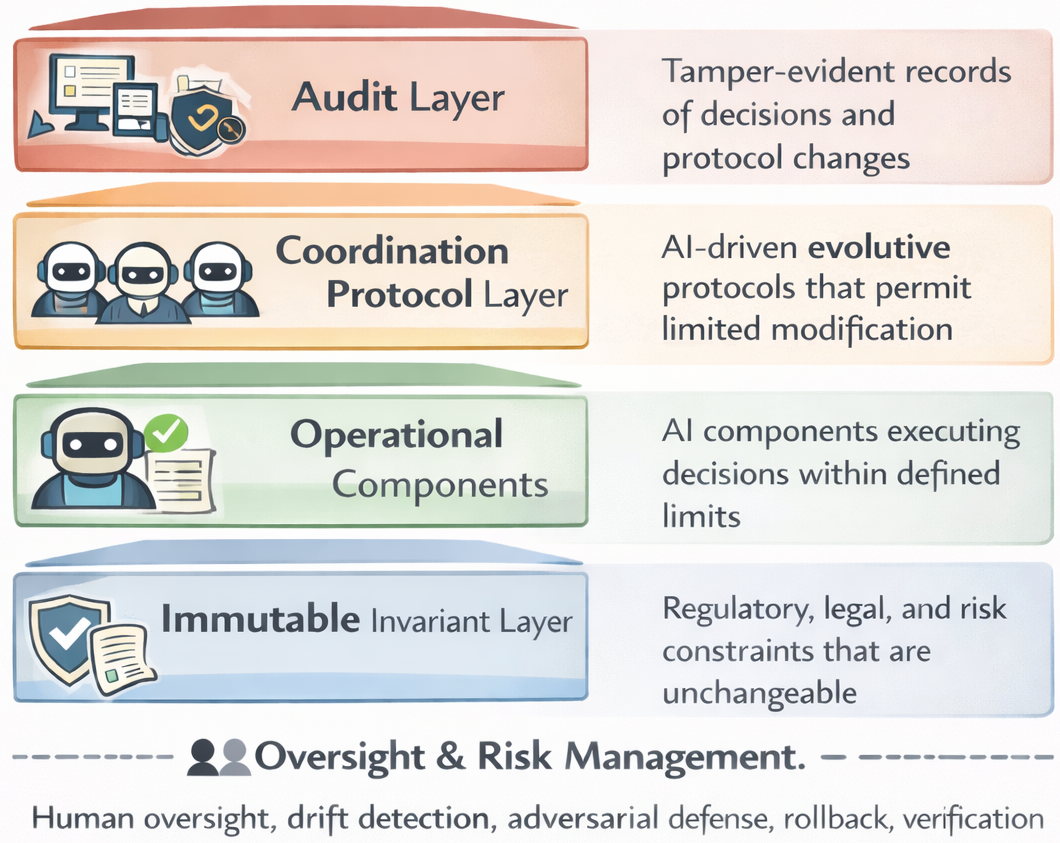}
    \caption{Layered governance architecture for production deployment. From bottom to top, the system is structured into an immutable invariant layer (regulatory, legal, and risk constraints), an operational layer executing decisions within those constraints, an evolutive coordination protocol layer permitting bounded self-modification, and an audit layer providing tamper-evident records of decisions and protocol changes. Human oversight and risk-management functions act across layers, enforcing approval, monitoring, rollback, and continuous compliance.}
    \label{fig:layered-governance}
\end{figure}

To make the production-deployment requirements concrete, Figure \ref{fig:layered-governance} illustrates the layered governance architecture discussed in this subsection. The diagram highlights the separation between immutable constraints, evolutive coordination logic, operational execution, and audit infrastructure, together with the cross-cutting role of human oversight and risk management. Its purpose is to clarify how bounded self-modification can be permitted at specific layers while higher-level invariants and oversight mechanisms remain fixed and enforceable.

\subsection{Regulatory Engagement}

Any credible deployment of such coordination mechanisms demands early and sustained engagement with regulators. Institutions must be prepared to disclose architectural designs, invariants, and modification procedures well before production use. Initial deployments should be limited to non-critical applications, with comprehensive auditability built in from the outset. Modification authority should remain conservative and human-in-the-loop until extensive operational evidence is accumulated.

Finally, it must be emphasized that the single-shot feasibility study presented in this paper is insufficient for regulatory validation. Production deployment would require substantially stronger evidence, including long-horizon stability studies, adversarial evaluations, and formal verification of invariant preservation. Without these, recursive coordination protocols should be regarded as experimental, not operational.

\section{Limitations}
\label{sec:limitations}

This section states, without mitigation rhetoric, the principal limitations of the experiment, the methodological constraints that follow from them, and the narrow bounds within which the results should be interpreted.

\subsection{Experimental design limitations}

The study’s single-shot design and single-step modification rule are the dominant constraints on what the experiment can and cannot show. Because each coordination protocol was executed exactly once on one fixed proposal set, the results are existence proofs—demonstrating that a particular outcome can occur under specified inputs and procedures but they do not support statistical inference. No confidence intervals, hypothesis tests, or estimates of variability are possible. Consequently, it is unknown whether the observed $50\%$ relative increase in coverage from \acs{SECP} v1.0 to \acs{SECP} v2.0 is representative, exceptional, or sensitive to random fluctuations in model outputs.

Closely related, only a single protocol modification iteration was performed (v1.0 → v2.0). That design choice precludes any analysis of long run dynamics: we cannot determine whether further iterations would improve coverage, converge to a fixed protocol, oscillate, or degrade system behavior. Claims are therefore limited to architectural capacity for a one-step, governed modification. No statements about multi-step recursion, stability, or optimality are supported.

The fixed proposal set (six Byzantine consensus designs) and small system scale (six modules, six proposals) further limit generalization. The proposals were selected to elicit disagreement; they do not constitute a representative sample of all possible designs or domains. Coordination dynamics and computational burdens may change qualitatively with larger numbers of modules or proposals, different problem domains (e.g., resource allocation, policy selection), or proposals exhibiting different technical trade-offs.

\subsection{System architecture and evaluator limitations}

Decision modules were instantiated with contemporary \acp{LLM}. Treating \acp{LLM} as technical evaluators introduces well-documented validity risks: hallucinations, incomplete formal reasoning compared with mechanized proof assistants, prompt sensitivity, and potential inconsistency across evaluations. The experiment reduced some surface risks (structured questionnaires, independent hard-constraint checks, and multiple perspectives), but these measures do not substitute for mechanical proof checking or independent expert adjudication. Thus the experiment evaluates coordination behavior \emph{conditional on} module outputs rather than providing strong evidence that module outputs are correct.

No adversarial testing was performed. All modules operated within the cooperative design intent. The experiment does not probe strategic manipulation, collusion, or attempts to game the modification process. Consequently, the security, manipulation/resistance, and Byzantine robustness of the coordination mechanisms remain unknown and are likely insufficient without explicit defenses.

Human operators and oversight mechanisms were not included in the empirical loop. Real-world governance requires human review, judgment, and intervention. The absence of human–\ac{AI} interaction experiments leaves open important questions about usability, accountability, and effective supervisory control.

\subsection{Evaluation methodology limitations}

Coverage, defined as the count of accepted proposals, was the primary metric. Coverage is a coarse, unweighted integer that captures one aspect of coordination outcome but omits critical dimensions: decision quality, correctness, deliberation efficiency, computational cost, robustness to strategic behavior, and explanation fidelity. Without ground-truth labels or external benchmarks for “\emph{optimal}” decisions, the experiment cannot measure accuracy, precision, recall, or other standard performance statistics. The binary Accept/Reject outcome further compresses information: it conceals degrees of consensus, near-miss cases, and the structure of unresolved objections. These choices simplify interpretation but substantially limit what can be concluded about the substantive quality of decisions.

\subsection{Theoretical and formal limitations}

Invariant preservation and other safety properties were checked empirically through logging and external inspection rather than proven formally. There is no machine-checked guarantee that an arbitrary permitted modification would preserve the declared invariants. Likewise, the study provides no theoretical analysis of convergence or complexity growth under repeated modification: there are no theorems, sufficient conditions, or worst-case bounds that would predict long-term behavior. Without formal proofs or complexity bounds, protocol modifications could in principle produce unbounded complexity, reduced explainability, or emergent behaviors that violate safety requirements in untested scenarios.

\subsection{Scope and interpretation boundaries}

Given these limitations, the correct interpretation of the results is narrow and explicit. The experiment demonstrates that a governed, bounded self-modification architecture can be implemented and can change measurable outcomes in a controlled run. It does not demonstrate that such a system is reliable, secure, or ready for production. It does not demonstrate statistical regularities, convergence, long-term stability, adversarial robustness, scalability to production sizes, or formal safety guarantees. Any claim beyond the narrow statement, \emph{``a single governed modification iteration changed coverage in this execution''}, is unsupported by the evidence presented.

\subsection{Recommended next steps}

To move from feasibility to practical validation, the following are necessary: 
\begin{itemize}
    \item Repeated, randomized trials to quantify variability
    \item Multi-iteration studies to examine dynamics and convergence
    \item Adversarial evaluations to surface manipulation vectors
    \item Human-in-the-loop experiments to assess oversight and accountability
    \item Integration with formal proof assistants for constraint verification
    \item Theoretical work to bound complexity growth and establish invariant-preservation proofs
\end{itemize}
Until some subset of these steps is completed, the architecture should be treated as experimental infrastructure, not production-ready governance.

\section{Conclusion}

\subsection{Summary of contributions}

This paper reports a focused systems feasibility study of bounded, governed self-modification for coordination protocols. In a controlled proof-of-concept experiment (six decision modules evaluating six Byzantine consensus proposals) we demonstrated that (i) contemporary \ac{AI} models can synthesize non-scalar coordination rules de novo and execute a validated single modification step; and (ii) this modification produced a measurable change in behaviour (coverage increased from two to three accepted proposals) while maintaining the experiment's declared invariants. We also characterized a persistent trade-off between protocol coverage (how many proposals a protocol will accept) and decision autonomy (the ability of individual evaluators to assert non-compensable objections). Finally, we sketched architectural patterns—immutable invariants, bounded modification spaces, explicit validation, and audit trails—that make such bounded self-modification governable in principle.

\subsection{Limitations and interpretation}

The empirical claims are narrow and intentionally circumscribed. The experiment was single-shot and single-iteration: results are existence proofs (a particular outcome can occur under the specified inputs and procedures) and do not support statistical generalization, claims about typical performance, or conclusions about long-run dynamics. Decision modules were \acp{LLM}. Their assessments are imperfect and not a substitute for mechanized proof checking or independent expert adjudication. No adversarial testing was performed, no human-in-the-loop governance was evaluated empirically, and system scale was small. Most critically, invariant preservation was validated by inspection and tests in bounded scenarios rather than by formal, machine-checked proofs. These limitations mean the work establishes architectural feasibility, not reliability, robustness, or production readiness.

\subsection{Practical implications for governance}

For regulated domains—finance being a primary example, the results imply that coordination protocols deserve explicit governance design. Non-scalar coordination can limit deadlock while preserving meaningful evaluative rights, and bounded self-modification can adjust protocol behavior without wholesale redesign. That said, practical adoption requires conservative governance: human approval for substantive changes, continuous monitoring, rollback capability, adversarial hardening, and formal verification of safety/critical invariants. Until those components are in place, recursive modification mechanisms should be treated as experimental governance tools rather than operational policy.

\subsection{Future research directions}

The highest-priority next steps are empirical, adversarial, and formal work that closes the gap between feasibility and deployment. In particular:
\begin{enumerate}
    \item Run multi-iteration experiments (multiple, randomized modification cycles) to observe dynamics, variability, and potential convergence or failure modes
    \item Develop and test adversarial scenarios (strategic manipulation, collusion, Byzantine coordination) and strengthen defensive mechanisms
    \item Integrate mechanized proof tools for machine-checked invariant verification and compare their outputs with \ac{LLM} assessments
    \item Scale experiments across larger module populations and diverse proposal spaces to assess computational and coordination bottlenecks
    \item Evaluate human-in-the-loop governance designs to measure oversight effectiveness, usability, and accountability
\end{enumerate}

\subsection{Final assessment}

The study establishes that bounded, governed protocol self-modification is technically implementable with current \ac{AI} systems and that such mechanisms can change coordination outcomes in measurable ways while preserving declared invariants in controlled runs. That is a necessary but far from sufficient condition for deployment. Substantial additional evidence repeated trials, adversarial validation, formal verification, human is required before these architectures can be considered safe, reliable, or fit for production in regulated settings. This work should therefore be read as a narrow architectural demonstration that motivates, but does not justify, further development and validation.

\section*{Acknowledgments}

The author thanks the AI Lab team at Grupo Santander for discussions and feedback, reviewers for constructive critique that improved the paper's technical precision and honesty about limitations, and the anonymous evaluators whose concerns about single-shot designs and N=1 limitations motivated explicit methodological justifications.

\appendix

\section{Decision module specifications}

This appendix specifies the six decision modules, the model instances used to realize them, and the narrow evaluation focus assigned to each. The descriptions emphasize what each module is intended to evaluate and the exact role it plays in the experiment; they do not assert superiority of any vendor model. All this information is shown in Table \ref{tab:decision-modules}

\begin{table}[ht]
    \centering
    \small
    \caption{Decision module specifications: model instantiation, remit, evaluation focus, and key assessment criteria.}
    \label{tab:decision-modules}
    \begin{tabular}{@{}p{2cm}p{2cm}p{3.3cm}p{4.2cm}p{3.3cm}@{}}
        \toprule
        \textbf{Module} & \textbf{Model} & \textbf{Expertise} & \textbf{Evaluation focus} & \textbf{Key criteria} \\
        \midrule
        Explorer & Sonnet 4.5 & Algorithmic novelty and design innovation & Identifies genuinely new techniques, previously unaddressed optimization dimensions, and mechanisms likely to influence future designs. & Novelty of approach; substantive conceptual advance; potential impact on design space. \\
        \hline
        Validator & Sonnet 4.5 & Formal correctness and proof quality & Examines safety and liveness arguments for logical completeness, explicit assumptions, and edge-case coverage. & Proof completeness; clarity of assumptions; coverage of edge cases and logical rigor. \\
        \hline
        Minimalist & Sonnet 4.5 & Simplicity and explainability & Assesses descriptive concision, state/minimality of protocol, and implementational transparency. & Conceptual simplicity; minimal state and transitions; operational clarity and implementability. \\
        \hline
        Robustifier & GPT-4 & Fault tolerance and adversarial analysis & Probes protocol behaviour under the stated fault model and plausible Byzantine attack vectors. & Byzantine resilience; adversarial robustness; comprehensiveness of fault-model coverage. \\
        \hline
        Proofsmith & GPT-4 & Proof methodology and verification rigor & Evaluates appropriateness and scalability of proof techniques, inductive structure, and methodological soundness. & Suitability of proof techniques; clarity of invariants; methodological rigor and compositionality. \\
        \hline
        Economizer & GPT-4 & Computational and communication efficiency & Reviews message complexity bounds, per-node computation, and practical overheads affecting deployability. & Message-efficiency (asymptotic and practical); computational overhead; operational cost and implementability. \\
        \bottomrule
    \end{tabular}
\end{table}

\subsection*{Rationale for heterogeneous composition}

Using both Claude Sonnet 4.5 and GPT-4 reflects a multi-vendor deployment pattern common in production settings and provides architectural diversity in evaluation perspectives. This choice is pragmatic. It exposes the coordination layer to heterogeneous reasoning biases. No claim is made that one model family is intrinsically superior. The modules function as distinct evaluative lenses for the coordination protocols.

\section{Deterministic scoring schema}              
\label{sec:appendixb}

To reduce subjectivity in numerical assessments, modules answer standardized, checkable questions and map responses deterministically to scores. The scoring schema is designed to be reproducible and auditable, replacing arbitrary numeric judgments with traceable aggregations of factual answers.

\subsection*{Questionnaire design}

Questions are grouped into categories aligned with module expertise. Each question targets a concrete, verifiable property of the proposal (for example, whether a safety argument explicitly addresses message reordering). Questions are answered as \emph{Yes}, \emph{Partial}, or \emph{No}, mapped to numeric values $1.0$, $0.5$, and $0.0$ respectively.

The principal categories are correctness, efficiency, simplicity, robustness, and innovation. Correctness questions address completeness of safety and liveness arguments. Efficiency questions examine message and computational complexity. Simplicity questions assess descriptive clarity and implementation complexity. Robustness questions target fault-model coverage and adversarial behavior. Innovation questions capture substantive novelty relative to existing approaches.

\subsection*{Score computation}

Each module $m$ is assigned a set of relevant categories $\mathcal{C}_m$ and associated weights $w_c$, with $\sum_{c \in \mathcal{C}_m} w_c = 1$. For proposal $p$, the module score is computed as:
\[
\mathrm{S}_m(p) = \sum_{c \in \mathcal{C}_m} w_c \cdot \frac{1}{|Q_c|} \sum_{q \in Q_c} \tilde{y}_m(q,p),
\]
where $Q_c$ is the set of questions in category $c$ and $\tilde{y}_m(q,p) \in \{1.0, 0.5, 0.0\}$.

This mapping guarantees that, for a fixed proposal and question set, the resulting score is deterministic and fully traceable to individual responses.

\subsection*{Verification and limitations}

Deterministic scoring improves reproducibility and auditability by linking every numerical score to explicit answers. It does not eliminate evaluator error: incorrect answers remain possible. Scores should therefore be interpreted as outputs of a transparent procedure, not as definitive judgments of correctness.

\section{Audit trail structure}

This appendix describes the logging and audit infrastructure supporting transparency and accountability.

\subsection*{Contents of the audit record}

For each evaluation, the system records protocol identifiers and versions, coordination rules and parameters, declared invariants, timestamps, and execution context. Proposal artifacts and results of independent hard-constraint checks are stored verbatim.

Module-level records include question responses, computed scores, textual objections, and support indicators across deliberation rounds. For deliberative protocols, logs capture the sequence of exchanges, support evolution, number of rounds executed, and explicit termination conditions.

Final decision records contain the Accept or Reject outcome, the protocol version applied, and a rationale trace linking the outcome to specific module inputs and coordination steps.

\subsection*{Audit capabilities and guarantees}

The audit trail supports reconstruction of decisions, identification of objections and their sources, verification that coordination rules were followed, and documentation of protocol modifications with associated rationales. Logs are write-once and versioned, enabling tamper detection and rollback to prior protocol versions.

\subsection*{Practical caveats}

Comprehensive logging enhances transparency but does not guarantee correctness. An auditable incorrect decision remains incorrect; logs only make errors inspectable. Effective governance therefore requires audit infrastructure to be complemented by formal verification, expert review, and clearly defined oversight authority.

\bibliographystyle{ieeetr}
\bibliography{references}

\end{document}